\documentclass[preprint,trackchanges]{aastex62}
\received{}
\revised{}
\accepted{}
\submitjournal{ApJ}

\begin{document}

\title{OBSERVATIONAL EVIDENCE OF MAGNETIC RECONNECTION ASSOCIATED WITH MAGNETIC FLUX CANCELLATION}

\correspondingauthor{Bo Yang}
\email{boyang@ynao.ac.cn}

\author{Bo Yang}
\affil{ Yunnan Observatories, Chinese Academy of Sciences, 396 Yangfangwang, Guandu District, Kunming, 650216, P. R. China}
\affiliation{Center for Astronomical Mega-Science, Chinese Academy of Sciences, 20A Datun Road, Chaoyang District, Beijing, 100012, P. R. China}
\affiliation {Key Laboratory of Solar Activity, National Astronomical Observatories of Chinese Academy of Science, Beijing 100012, China }

\author{Jiayan Yang}
\affil{ Yunnan Observatories, Chinese Academy of Sciences, 396 Yangfangwang, Guandu District, Kunming, 650216, P. R. China}
\affiliation{Center for Astronomical Mega-Science, Chinese Academy of Sciences, 20A Datun Road, Chaoyang District, Beijing, 100012, P. R. China}

\author{Yi Bi}
\affil{ Yunnan Observatories, Chinese Academy of Sciences, 396 Yangfangwang, Guandu District, Kunming, 650216, P. R. China}
\affiliation{Center for Astronomical Mega-Science, Chinese Academy of Sciences, 20A Datun Road, Chaoyang District, Beijing, 100012, P. R. China}

\author{Junchao Hong}
\affil{ Yunnan Observatories, Chinese Academy of Sciences, 396 Yangfangwang, Guandu District, Kunming, 650216, P. R. China}
\affiliation{Center for Astronomical Mega-Science, Chinese Academy of Sciences, 20A Datun Road, Chaoyang District, Beijing, 100012, P. R. China}

\author{Haidong Li}
\affil{ Yunnan Observatories, Chinese Academy of Sciences, 396 Yangfangwang, Guandu District, Kunming, 650216, P. R. China}
\affiliation{Center for Astronomical Mega-Science, Chinese Academy of Sciences, 20A Datun Road, Chaoyang District, Beijing, 100012, P. R. China}

\author{Zhe Xu}
\affil{ Yunnan Observatories, Chinese Academy of Sciences, 396 Yangfangwang, Guandu District, Kunming, 650216, P. R. China}
\affiliation{Center for Astronomical Mega-Science, Chinese Academy of Sciences, 20A Datun Road, Chaoyang District, Beijing, 100012, P. R. China}
\affiliation{University of Chinese Academy of Sciences, 19A Yuquan Road, Shijingshan District, Beijing 100049, China}

\author{Hechao Chen}
\affil{ Yunnan Observatories, Chinese Academy of Sciences, 396 Yangfangwang, Guandu District, Kunming, 650216, P. R. China}
\affiliation{Center for Astronomical Mega-Science, Chinese Academy of Sciences, 20A Datun Road, Chaoyang District, Beijing, 100012, P. R. China}
\affiliation{University of Chinese Academy of Sciences, 19A Yuquan Road, Shijingshan District, Beijing 100049, China}


\begin{abstract}
Using high spatial and temporal data from the \emph{Solar Dynamics Observatory} (\emph{SDO})
and the \emph{Interface Region Imaging Spectrograph} (\emph{IRIS}),
several observational signatures of magnetic reconnection in the course of magnetic flux cancellation are presented,
including two loop-loop interaction processes, multiple plasma blob ejections, and a sheet-like structure
that appeared above the flux cancellation sites with a Y-shaped and an inverted Y-shaped ends.
The \emph{IRIS} 1400 \AA\ observations show that the plasma blobs were ejected from
the tip of the Y-shaped ends of the sheet-like structure.
Obvious photospheric magnetic flux cancellation occurred after the first loop-loop interaction
and continued until the end of the observation.
Complemented by the nonlinear force-free field extrapolation,
we found that two sets of magnetic field lines, which reveal an X-shaped configuration,
align well with the interacted coronal loops.  Moreover, a magnetic null point is found to be
situated at about $0.9$ Mm height right above the flux cancellation sites
and located between the two sets of magnetic field lines. These results suggest that
the flux cancellation might be a result of submergence of magnetic field lines following magnetic reconnection
that occurs in the lower atmosphere of the Sun, and the ejected plasma blobs should be plasmoids created
in the sheet-like structure due to the tearing-mode instability.
This observation reveals detailed magnetic field
structure and dynamic process above the flux cancellation sites and will
help us to understand magnetic reconnection in the lower atmosphere of the Sun.	

\end{abstract}


\section{Introduction} \label{sec:intro}
Magnetic reconnection is a process by which magnetic field lines with antiparallel components
are brought together in a current sheet or at a magnetic null point, where they break up and reconnect
to form new magnetic field lines\citep{pri00,yama10}. During this process
magnetic energy is thereby converted into plasma kinetic and thermal energy.
It is widely accepted that magnetic reconnection is the cause of various types of solar activities,
such as solar flares\citep{shi96a}, coronal mass ejections\citep[CMEs;][]{lf00},
filament eruptions\citep{cs00,shen12,zhou17}, jets\citep{shi96b,jiang13}, explosive events\citep{innes97},
and coronal bright points\citep{pri94}. To date, many signatures that are probably related to magnetic reconnection
have been reported, including hot cusp-shaped structures\citep{tsu92},
loop-top hard X-ray sources\citep{mas94,sui03}, reconnection inflows\citep{yoko01,li09,su13,sun15,yang15}
and outflows\citep{asai04,sav10,liu13,chen16}, current sheets\citep{web03,lin05,liu10,xue16,yan18},
plasmoid ejections\citep{shi95,nish10,taka12}, loop-loop interactions\citep{sak96,li14}, and drifting pulsating structures observed
in radio waves\citep{kilm00,ning07}. Through decades of observations, a lot of
evidence for the reconnection scenario has been obtained. However, most evidence was indirect and was detected in the solar corona,
and direct observational evidence that characterizes the reconnection in the lower atmosphere has been poorly reported.

Magnetic flux cancellation, which observationally describes the mutual disappearance of converging magnetic patches
of opposite polarities in the photospheric longitudinal magnetograms\citep{liv85,mar85},
is considered to be evidence of magnetic reconnection occurring in the lower atmosphere of the Sun\citep{pri94}.
A ``U-loop emergence" scenario and an ``$\Omega$-loop submergence" scenario were proposed by \citet{zwaan87} to
account for magnetic flux cancellation. Two unconnected magnetic patches of opposite polarities could build up
connection by magnetic reconnection during flux cancellation\citep{wang93}, and whether a ``U-loop emergence"
scenario or an ``$\Omega$-loop submergence" scenario could contribute to flux cancellation depends on the
height that reconnection is initiated. The ``U-loop emergence"  will be dominant during the cancellation when
magnetic reconnection takes place below the photosphere. On the contrary, the ``$\Omega$-loop submergence"
will be dominant when magnetic reconnection occurs above the photosphere. By investigating the evolution of
the photospheric and chromospheric magnetograms simultaneously, \citet{har99} proposed strong evidence that
suggests an ``$\Omega$-loop submergence" scenario at flux cancellation sites. Transverse magnetic field and
Doppler velocity field around flux cancellation sites are usually utilized to study flux cancellation
events. During and after flux cancellation, it is usually found that horizontal field at the flux cancellation
sites enhanced significantly\citep{wang93,yang16}. However, \citet{wang93} implied that the change of the
horizontal field at flux cancellation sites could not fit the quite popular view of interpreting flux cancellation
that mentioned above. They put forward that the association of flares to flux cancellation seems to represent
the coupling of a slow reconnection in the lower atmosphere to a fast reconnection in the upper atmosphere.
\citet{chae04} and \citet{iida10} verified that both red shifts and strong horizontal field at flux cancellation
sites support the ``$\Omega$-loop submergence" scenario. \citet{zhang09} reported extremely large Doppler blue-shifts
at flux cancellation sites and interpreted the cancellation as a ``U-loop emergence".
\citet{kubo07} found that there are both blue and red shifts at flux cancellation sites, indicating that
magnetic reconnection between the converging magnetic patches occurs at multiple locations with different heights.
Nevertheless, \citet{yang09} investigated the emerged dipoles in a coronal hole and found that the submergence of the
emerged original loops can also lead to flux cancellation. Therefore, to understand the physical nature of flux cancellation,
the detailed magnetic structures above the flux cancellation sites in the upper atmosphere need to be investigated in detail.

In this paper, with high resolution observations acquired by the \emph{Solar Dynamics Observatory}
\citep[\emph{SDO};][]{pes12} and the \emph{Interface Region Imaging Spectrograph} \citep[\emph{IRIS};][]{depont14},
we present clear and direct observational evidence showing magnetic reconnection
associated with photospheric magnetic flux cancellation.
This is an exemplary event with which to show in detail the relationship between magnetic reconnection and photospheric magnetic flux cancellation.
In Section 2, we describe the detailed observations and methods that we used. The results are shown in Section 3.
The conclusion and the discussion are given in Section 4.

\section{Observations and Methods}
The detailed reconnection process associated with magnetic flux cancellation on 2015 January 9 was captured by
the Atmospheric Imaging Assembly \citep[AIA;][]{lem12} and the Helioseismic and Magnetic Imager \citep[HMI;][]{sch12}
on board the \emph{SDO}. The AIA instrument observes full-disk images of the Sun in 10 ultraviolet (UV) and extreme
ultraviolet (EUV) wavelengths with a spatial resolution of 1.$\arcsec$5 (0.$\arcsec$6 pixel$^{-1}$) and a high cadence
of up to 12 s. In this study, we mainly used the Level 1.5 images observed in 304 \AA \ (\ion{He}{2}, 0.05 MK),
171 \AA \ (\ion{Fe}{9}, 0.6 MK), 94 \AA \ (\ion{Fe}{18}, 7 MK), and 1600 \AA \ (\ion{C}{4} + cont., 0.01 MK).
HMI measures the full-disk continuum intensity images and line of sight (LOS) magnetic field for the \ion{Fe}{1}
absorption line at 6173 \AA\, with a spatial sampling of 0.$\arcsec$5 pixel$^{-1}$ and a cadence of 45 s.
The AIA data used in this study were taken between 2015 January 9 19:30 UT and 21:30 UT, and the HMI data
were taken between 2015 January 9 18:00 UT and 22:00 UT. This event was also observed by \emph{IRIS} slit jaw
imager (SJI) in 1400 \AA\ during two periods (19:03-20:00 UT; 20:40-21:32 UT). The time cadence
and the spatial resolution of the SJIs are 9 s and  0.$\arcsec$332 pixel$^{-1}$, respectively.
Using full-disk soft X-ray (SXR) images from the X-Ray Telescope (XRT) aboard the \emph{Hinode} satellite \citep{kos07},
the associated coronal structures were also examined.
All images were then aligned by differentially rotating to a reference time (20:40 UT on 2015 January 9).

In addition, continuous photospheric vector field \citep{tur10}, which has a pixel scale of about 0.$\arcsec$5 pixel$^{-1}$
and a cadence of 12 minutes, in the so called HMI Active Region Patches (HARPs) region is also provided by HMI.
The Very Fast Inversion of the Stokes Vector algorithm \citep{bor11} is utilized to compute the vector field data, and
the Minimum Energy method \citep{met1994,met2006,lek2009} is used to resolve the remaining 180\arcdeg\ azimuth ambiguity.
In order to remove the projection effect, the HARP vector field data are remapped to a Lambert Cylindrical Equal-area (CEA)
projection and then transformed into standard heliographic spherical coordinates. To obtain the magnetic field topology of the
flux cancellation event, we carried out a nonlinear force-free magnetic field (NLFFF) extrapolation to reconstruct the coronal fields.
To perform the NLFFF extrapolation, the "weighted optimization" method \citep{whe2000,wie04} is used. Before the extrapolation,
a preprocessing procedure, which drives the observed non-force-free data towards suitable boundary conditions for
a force-free extrapolation \citep{wie06}, is applied to the bottom boundary vector data.

\section{Results}
\subsection{Cancellation of Photospheric Magnetic Field}
On 2015 January 9, AR NOAA 12257 was located at about N$5\degr$W$29\degr$ with a $\beta$ magnetic configuration.
As shown in Figure 1({\it a}), the magnetic flux cancellation region of interest is enclosed by a red rectangle, and
the detailed magnetic flux cancellation process is shown in the zoomed view in panels({\it b} \sbond {\it f}).
The cancelling magnetic flux patches ``p" and ``n1" existed from the beginning of the observations, and a transverse field,
which was emanated  from p and connected to n1, indicates that there was
a connectivity between p and n1(panel ({\it b})).
Note that flux emergence happened before  20:00 UT (panels({\it b} \sbond {\it c})).
The positive flux patches of the emerged flux were mixed with p,
while its negative flux patches were composed of ``n2" and ``n3".
In particular, during its emerging process, n3 moved toward and merged with n1. As a result, the flux density and the area of p and n1 increased,
although the flux cancellation occurring between p and n1. A remarkable decrease of the flux density and the area of n1 was observed
from 20:00 UT to 21:40 UT (panels({\it d} \sbond {\it g})), and the unsigned negative flux was dropped by $8.0\times10^{19} \, Mx$,
corresponding to an approximate flux cancellation rate of $ 4.8\times10^{19} \, Mx\, h^{-1}$.
At the end of the observations, n1 almost disappeared (panel({\it f})).
Different from many flux cancellation events observed before\citep{wang93,yang16},
the change of  the transverse field was not obvious,
and the flux cancellation was accompanied by the flux emergence at the same region.
Therefore, it is difficult to confirm which mechanism could account for the flux cancellation.
Investigating the coronal structures and activities above the flux cancellation sites
may shed light on the understanding of the physical nature of the flux cancellation.

\subsection{The First Loop-Loop interaction Process}
Scrutinizing the observations from \emph{SDO}/AIA and \emph{IRIS}, it is found that two loop-loop interaction processes
and two plasma blob ejection processes were closely related to the flux cancellation.
The first loop-loop interaction process is displayed in Figure 2 (see also the animation, loop-loop1.mpeg).
Just prior to the interaction, at about 19:40 UT, an \emph{IRIS} 1400 \AA\ image
shows the general appearance of the two sets of interacted loops, ``L1" and ``L2" (panel({\it a})).
Remarkably, as shown by the contoured HMI magnetogram, L1 connected the
positive flux patch ``p1" to a negative flux patch n1,
whereas L2 connected the positive flux patch p to a negative flux patch n2.
Thus, the adjacent endpoints of L1 and L2 were co-spatial with the cancelling flux patches p and n1.
By about 19:43 UT, the loop-loop interaction started, and a set of rising loops, ``L3",
which connected p1 to n2, was formed  (panels({\it b} \sbond {\it c})).
At the same time, four footpoint brightenings, which were exactly coincident with the footpoints of L1 and L2,
appeared (panel({\it c} and {\it f})). Note that L1 could not be detected by the AIA observations before the interaction.
However, after the interaction, L1 and L3 were clearly presented by the AIA 94 \AA\ and SXR images (panels({\it d} \sbond {\it e})),
implying that L1 and L3 might be heated during the interaction and the connectivity of L1 were partially changed.
Furthermore, a set of loops, ``L4" , which connected p to n1,
was also observed by the AIA 94 \AA\ and SXR observations.
These observations indicate that magnetic reconnection may take place between L1 and L2.
The reconnection changed the connectivity of L1 and L2, resulting in the formation of L3 and L4.

\subsection{Successive Ejection of Plasma Blobs from the Flux Cancellation Sites}
It is widely accepted that plasma blob ejections and magnetic flux cancellation are evidence of magnetic reconnection. Currently,
plasma blob ejections were frequently observed in different types of reconnection events; however, successive plasma blob ejections
associated with flux cancellation has rarely been observed. In our observations, from 20:25 UT to 20:40 UT, about half an hour after
the first loop-loop interaction, we found that a chain of plasma blobs were ejected from the flux cancellation sites successively.
The detailed ejection process is displayed  by the selected AIA 304 \AA\ images in Figure 3 (see also the animation, blobs1.mpeg).
Before the initiation of the plasma blob ejection at about 20:21 UT,
it is found that two adjacent bright streaks were rooted in the flux cancellation
patches p and n1, respectively (panel({\it a})). Those bright streaks may represent two sets of loops with opposite directions.
In particular, the bright streak rooted in n1 had the same connectivity as L1.
Hereafter, we call the loops, which connect p1 to n1, as L1.
As soon as those bright streaks approached to each other, the plasma blobs, as indicated by the arrows, were
ejected from the flux cancellation sites, propagated along L1 and finally stopped at the far ends of L1
(panels ({\it b} \sbond {\it d})).  The plasma blob ejections observed here are quite similar to that reported by \citet{zhang16}.
Generally, it is believed that those ejected blobs are formed by a tearing-mode instability occurring
in a current sheet structure\citep{furth63,shi01}. Thus, our observations may also imply that
magnetic reconnection and a tearing process may occur in a current sheet between
the two bright streaks.

\subsection{The Second Loop-Loop Interaction Process}
Immediately after the plasma blob ejection, an intense activity was followed by (see also the animation, loop-loop2.mpeg).
This activity is quite obvious in the AIA 304 \AA\ images in Figure 4({\it a} \sbond {\it c}). At about 20:40 UT, when the
plasma blob ejections stopped, a compact brightening appeared at the flux cancellation sites (panel({\it a})).
Simultaneously, relatively weak remote brightening appeared at the location corresponding to the negative footpoint of
L1. Subsequently, from 20:42 UT to 20:50 UT, mass flows, which originated from the compact brightening region,
spread along two arched trajectories in opposite directions (panels({\it b} \sbond {\it c})).
Careful inspecting the AIA 171 \AA\ difference image (panel({\it d})) found
that mass flows moved along the two arched trajectories 
and traced out the appearance of two sets of loops, L1 and ``L5".
Supplemented by the contoured HMI magnetogram, one can see that the adjacent ends of L1 and L5
are rooted in the cancelling flux patches p and n1 (panel ({\it d})), respectively. Moreover,
the negative footpoint of L5 rooted in a plage region (labeled as ``n").
These observational signatures may suggest that the compact brightening
and the plasma flows are the results of the interaction occurring between L1 and L5.

\subsection{ Successive ejection of Plasma Blobs from a Sheet-Like Structure Observed by IRIS}
At 21:03 UT, about 13 minutes after the second loop-loop interaction, it
is particularly remarkable that a sheet-like structure
with a Y-shaped and an inverted Y-shaped ends appeared
above the flux cancellation sites (Figure 5({\it a})). Afterwards, multiple plasma blobs
stemmed likely from the tip of the Y-shaped end and ejected successively along L1.
This is evidenced by the sequential \emph{IRIS} 1400 \AA\ images in Figure 5({\it a} \sbond {\it d}) (see also the animation, blobs2.mpeg).
The zoomed view (panel({\it e})) displays the morphology of the sheet-like structure more clearly.
This sheet-like structure, which is similar to the sheet-like structure
reported by \citet{singh12} and \citet{li16}, lasted about 8 minutes and finally disappeared at about 21:11 UT.
\citet{singh12} and \citet{li16} suggested that this structure should be a current sheet.
In our observations, however, there is no direct evidence to confirm that
this sheet-like structure is a current sheet apart from its morphology. Fortunately, the vector field data obtained by
HMI is conducive to extrapolate and reconstruct the coronal magnetic field over the flux cancellation region,
and is helpful for us to understand the event.

\subsection{Magnetic Topology of the Flux Cancellation Region}
With the aid of the HMI vector magnetograms, we carried out an NLFFF extrapolation to reconstruct the coronal magnetic
field of the flux cancellation region. Figure 5({\it g} \sbond {\it f})) show the consequence of the NLFFF extrapolation.
The red and blue lines, which traced from the photospheric flux patches p and n1, delineate the extrapolated coronal field lines.
It is evident that the red and blue field lines reveal an X-shaped configuration.
Previous theoretical and observational studies\citep{pri94,jiang17} suggested that
such a configuration should contain a magnetic null point, which is in favour of the reconnection.
Employing a trilinear null finding method\citep{hay07} to scan the NLFFF-modeled field,
we indeed find that a magnetic null point is located between the red and blue filed lines
(as indicated by the green arrows in Figure 5({\it f} \sbond {\it g})). The magnetic null point
is situated at $\sim 0.9$ Mm height right above the flux cancellation sites.
It separates the red and blue field lines into two distinct connections, one connects p1-n1,
and the other connects p-n. From Figure 4({\it d}) and Figure 5({\it g}), it is found that
the red and blue field lines match strikingly well with L1 and L5. Magnetic field near an X-type
null point would collapse and evolve to a field with a current sheet\citep{pri00}.
In our event, the X-shaped  magnetic filed configuration may imply that the magnetic null point is an X-type
null point. Moreover, the spatial location of the magnetic  null point and the observed sheet-like structure
is almost overlapping. Thus, our observations strongly suggest that the reconnection occurring between L1 and L5
was triggered at the magnetic null point, and the
magnetic field near the null point collapsed during the reconnection, resulting in the formation
of a current sheet.  Accordingly, we speculate that the sheet-like structure observed by the \emph{IRIS}
may represent a current sheet. A tearing-mode instability \citep{furth63,pri00}
may further develop in the sheet-like structure, creating the multiple plasma blobs.

\section{Conclusion and Discussion}
In this paper, we present two unambiguous loop-loop interaction processes
and two plasma blob ejection processes, which are closely related to magnetic flux cancelation in the same location.
The first loop-loop interaction took place between a set of pre-existing loops(L1) and a set of emerging small loops (L2).
Half an hour after the first loop-loop interaction, a chain of plasma blobs were
ejected from the flux cancellation sites and spread along L1.
Immediately after the plasma blobs ejection, the second loop-loop interaction initiated. Compact brightening resided at
the flux cancellation region and mass flows spread in opposite directions were observed.
The mass flows traced out the interacted loops L1 and another set of loops (L5). Following the second loop-loop interaction, \emph{IRIS} 1400 \AA\
images show that a sheet-like structure with a Y-shaped and an inverted Y-shaped ends appeared above the flux cancellation sites and
a chain of plasma blobs were ejected successively from the tip of the Y-shaped ends and moved along L1.
It is evident from HMI vertical magnetograms that obvious flux cancellation occurred after the first loop-loop interaction
and continued till the end of the observation. Supplemented by an NLFFF extrapolation,
two sets of coronal field lines, which align with L1 and L5 very well, are extrapolated. Moreover,
it is found that a magnetic null point is located between the two sets of coronal field lines. Based on the observations,
we suggest that the first loop-loop interaction may due to the magnetic reconnection between L1 and L2,
while the second loop-loop interaction may due to the magnetic reconnection between L1 and L5.
Furthermore, a tearing-mode instability might be further developed in the course of the interaction between L1 and L5 in a current sheet,
creating the ejected plasma blobs. Our observations not only provide evidence of a submergence of ``$\Omega$-loop "
following magnetic reconnection at the flux cancellation sites,
but also shed new light on magnetic reconnection in the lower atmosphere of the Sun.

Previous theoretical models have suggested that there should be magnetic null point and current sheet
around the flux cancellation sites in the upper atmosphere\citep{pri94,von2006}. 
Numerous observations mainly focused on the change of the velocity field and the transverse field around flux cancellation sites,
while direct observation of the detailed structure and the dynamic process above the flux cancellation sites  were extremely rare.
In our event, the loop-loop interactions should be the evidence of magnetic reconnection, the sheet-like structure
revealed by the \emph{IRIS} 1400 \AA\ images should be a current sheet resided above the flux cancellation sites.
Moreover, the extrapolated coronal field lines and the detected magnetic null point may further evidence that
magnetic reconnection would occur above the flux cancellation sites. These results are comparable with
the theoretical models of \citet{zwaan87} and \citet{pri94}, and reveal
detailed magnetic field structure above the flux cancellation sites.
According to the theoretical models of \citet{zwaan87} and \citet{pri94}, we can naturally explain our observations as follow:
as L1 contacts L2 or L1 contacts L5 in a magnetic null point, magnetic field near the
null point would collapse and evolve
to a field with a current sheet (as shown by the sheet-like structure in Figure 5({\it e})). Magnetic
reconnection between L1 and L2 or between L1 and L5 happened at the null point or inside the current sheet,
leading to the formation of a set of long loops that connects their far ends of the interacted loops
and a set of short loops (L4) that connects their adjacent ends. Caused by magnetic tension,
L4 further submerged and resulted in the flux cancellation.
The theoretical models of \citet{zwaan87} and \citet{pri94} are suitable for interpreting
the two loop-loop interactions and the associated flux cancellation.
However, the detailed dynamic processes above the flux cancellation sites,
for instance, the plasma blob ejections, need further investigation.

An important observational signature in our event is the ejection of plasma blobs.
As mentioned above, these plasma blobs were ejected from the tip of the Y-shaped ends of the sheet-like structure,
and the extrapolated coronal field lines reveal an X-shaped configuration containing a magnetic null point.
Accordingly, we inferred that the ejected plasma blobs should be plasmoids,
which are created by the tearing-mode instability occurring in the current sheet\citep{furth63,bha09}.
Previously, plasmoids are frequently observed in the coaxial bright rays
that appears in white light images in the wake of the CMEs\citep{lin05},
in the current sheet of solar flares\citep{taka12,kum13}, and in some jets \citep{singh12,zhang14,zhang16,zhang17}.
More recently, a detailed formation and evolution process of plasmoids was reported by \citet{li16}.
They found that the plasmoids appeared within the current sheets at the interfaces between an erupting filament
and nearby coronal loops and propagate bidirectionally along them, and then further along the
filament or the loops. In the lower atmosphere, continuous
ejections of plasmoids from the flux cancellation sites 
have seldom been observed directly. Through the numerical simulation method, \citet{ni15} simulated magnetic reconnection process
in partially ionized solar chromosphere and confirmed that fast magnetic reconnection mediated
by tearing-mode instability could be indeed triggered.
In particular, by analysing the \ion{Si}{4} line profiles obtained from the flux cancellation sites
of some small-scale reconnection events, \citet{innes15} suggested that a fast reconnection proceeding
via tearing-mode instability may play a central role in those small-scale reconnection events.
In the present case, the continuous ejection of plasma blobs above the flux cancellation sites
from the tip of the Y-shaped ends of the sheet-like structure are the direct observational evidence
that support the idea of \citet{innes15}. This observation displays
the detailed dynamic process above the flux cancellation sites, and has a significant physical implication for
the magnetic reconnection in the lower atmosphere of the Sun.

Before the first loop-loop interaction, we notice that there is connectivity between the cancelling flux patches p and n1,
and there is lack of velocity field information around the flux cancellation sites.
Moreover, new magnetic flux emerged beside the cancelling flux patches, and the emerged positive flux patches mixed with p,
while parts of its negative flux patches moved and merged with n1.
Thus, it is difficult to absolutely rule out the possibility that the submergence of original loops connecting p to n1
may also contribute to the magnetic flux cancellation. However,
it is clear from the time profile of flux changes (Figure 1({\it g})) that obvious flux cancellation
was observed after the first loop-loop interaction and continued till the end of the observation.
This time interval covers the second loop-loop interaction process and the two plasma blob ejection processes
(as indicated by the pink shadow in Figure 1({\it g})). Therefore, our observations support a causal relationship among
the loop-loop interactions, the plasma blob ejections, and the flux cancellation.

\acknowledgments
We thank the anonymous referee for useful comments and suggestions that have improved the quality of the manuscript.
We are very grateful to the AIA, HMI, \emph{IRIS}, and \emph{Hinode} teams for free access to data.
This work is supported by the Natural Science Foundation of China, under grants 11703084, 11633008,
11333007, 11503081, and 11503082, and the CAS ``Light of West China" Program,
and the Open Research Program of the Key Laboratory of Solar Activity of Chinese Academy of Sciences (KLSA201809),
and the CAS grant ``QYZDJ-SSW-SLH012",
and by the grant associated with the Project of the Group for Innovation of Yunnan Province.

\newpage
\begin{figure}
\epsscale{1.0}
\plotone{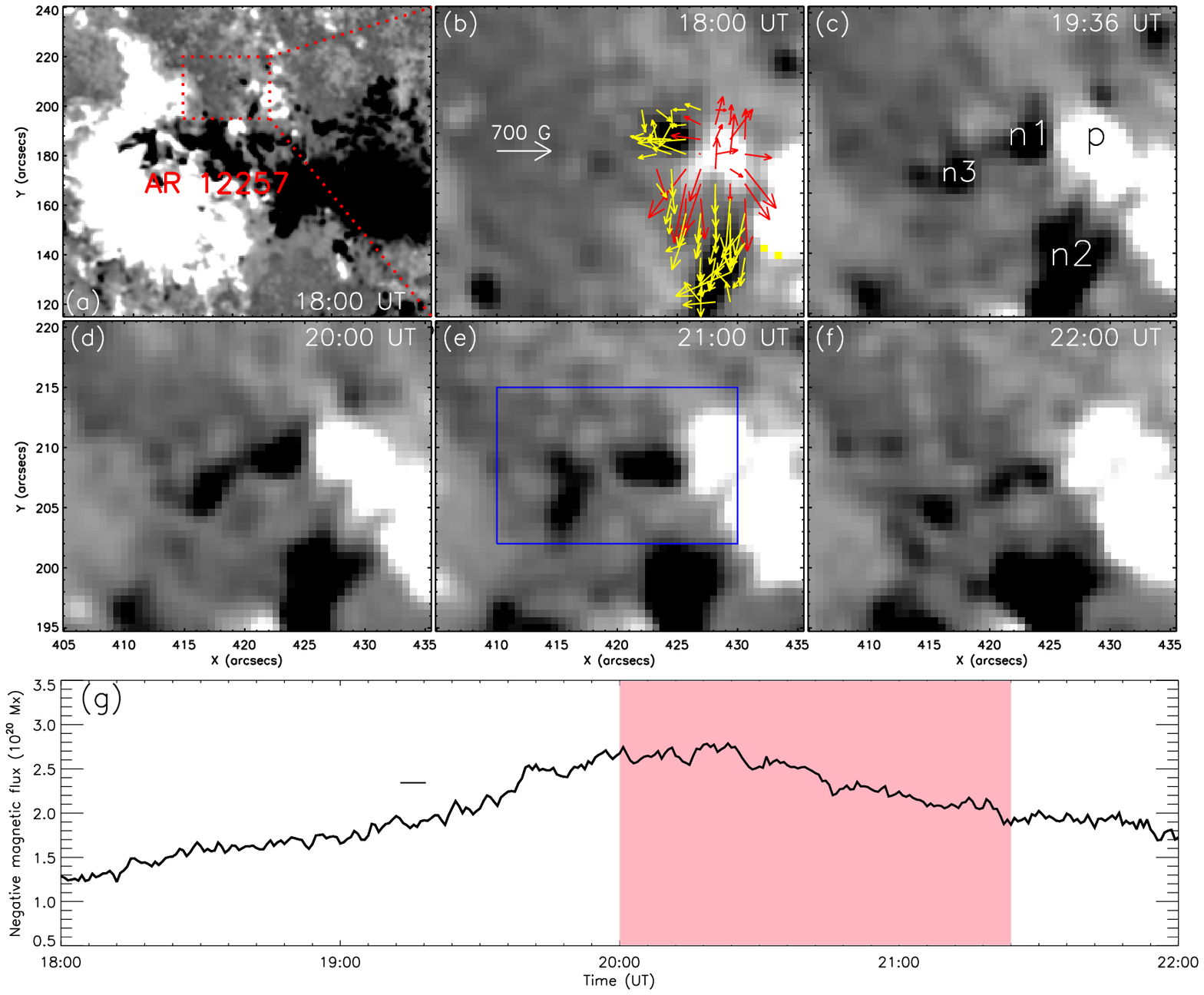}
\caption{\emph{SDO}/HMI vertical images displaying the general appearance of the NOAA AR 12257 at 18:00 UT on 2015 January 9 (panel ({\it a}))
and the cancellation of opposite polarities (panels ({\it b}\sbond{\it f})). Negative and positive magnetic flux patches are denoted as
``n1", ``n2", ``n3", and ``p", respectively.
Panel ({\it g}) showing the changes in negative magnetic flux in the blue box in panel ({\it e}).  In panel ({\it b}),
the transverse fields are overplotted as red and yellow arrows, which originate from a positive and negative longitudinal
field, respectively. The field of view (FOV) of panels ({\it b})\sbond({\it f}) is outlined by the red rectangle in panel ({\it a}).
The vertical pink shadow in panel ({\it g}) denotes the time when the flux cancellation occurs obviously.
}
\end{figure}

\begin{figure}
\epsscale{1.}
\plotone{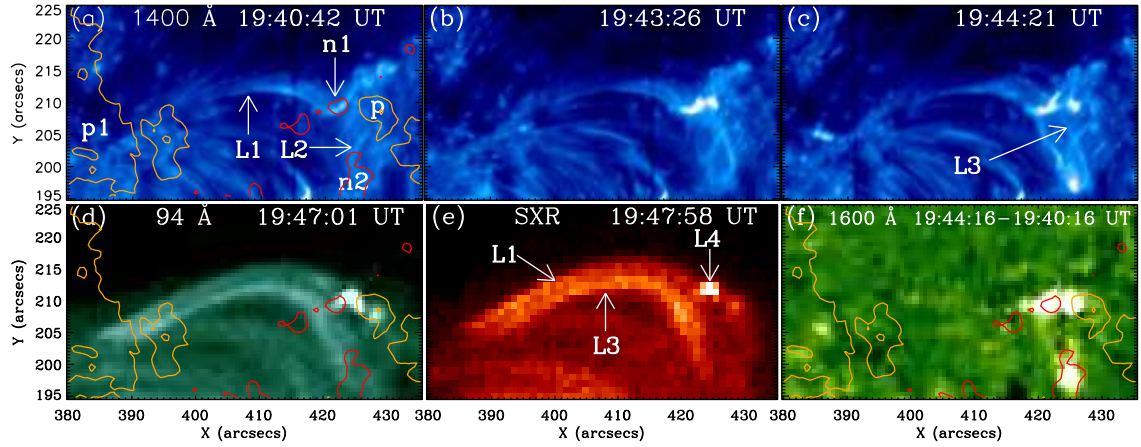}
\caption{\emph{IRIS} 1400 \AA \ SJI images ({\it a})\sbond({\it c}) showing the first loop-loop interaction process
that took place between two sets of loops,``L1" and ``L2". AIA 94 \AA\ image ({\it d}) and \emph{Hinode}/XRT SXR image ({\it e})
showing the coronal connectivity after the interaction. ``L3" and ``L4" denote the newly formed loops. Brightenings located
at the footpoints of L1 and L2 are distinctly showed in the AIA 1600 \AA\ difference image ({\it f)}. The orange and red contours
overplotted on panels ({\it a}), ({\it d}), and ({\it f}) represent the intensity contours of the positive and negative magnetic fields,
with contour levels of 200 G and -100 G, respectively. Likewise, ``n1", ``n2", ``n3", ``p", and ``p1" denote negative
and positive magnetic flux patches. An animation of panels ({\it a}\sbond{\it c}) and ({\it d}) is available. The animation is 2s in duration, covering
19:39:25 UT to 19:49:49 UT.
(An animation of this figure is available.)}
\end{figure}

\begin{figure}
\epsscale{0.6}
\plotone{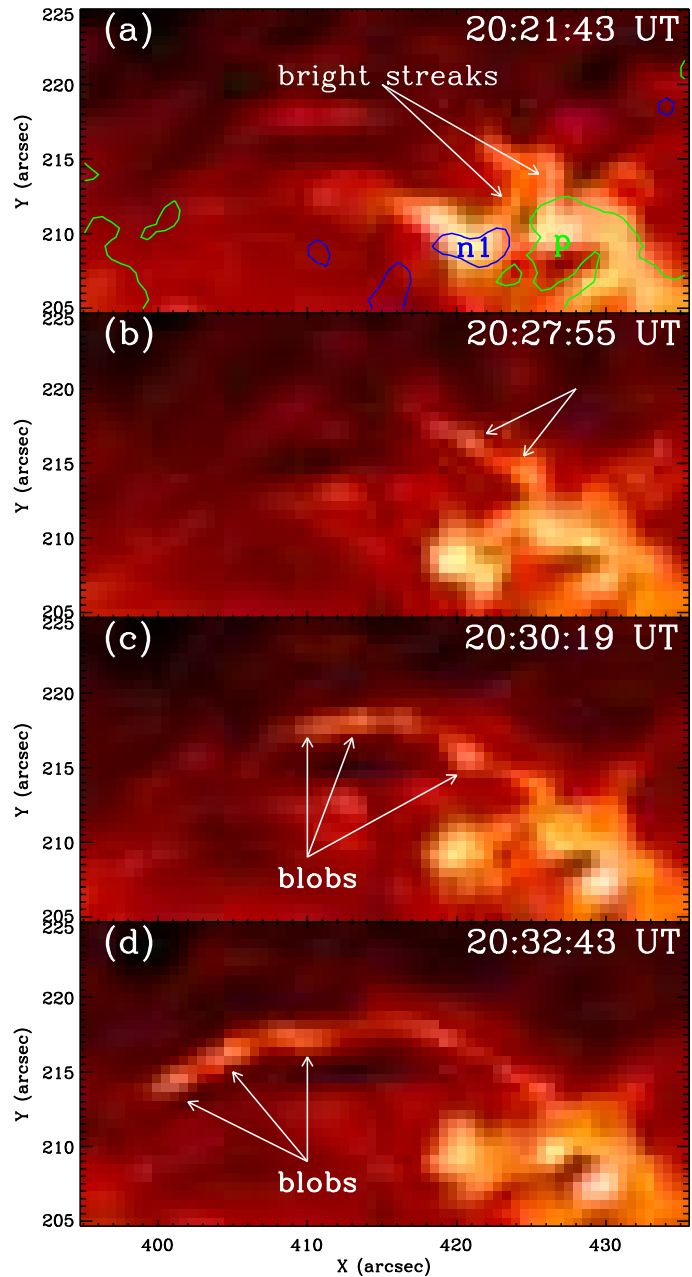}
\caption{AIA 304 \AA\ images ({\it a})\sbond({\it d}) present the successive ejection of plasma blobs
(as indicated by the arrows in panels ({\it b} \sbond {\it d})) from the flux cancellation sites.
Two bright streaks, which may imply two sets of interacted loops, are clearly seen in panel ({\it a}).
Iso-Gauss contours of  $\pm 100 G$ are superposed by green and blue lines on panel ({\it a}).
An animation of this figure is available. The animation is 1s in duration, covering
20:19:07 UT to 20:36:55 UT.
(An animation of this figure is available.)}
\end{figure}

\begin{figure}
\epsscale{0.6}
\plotone{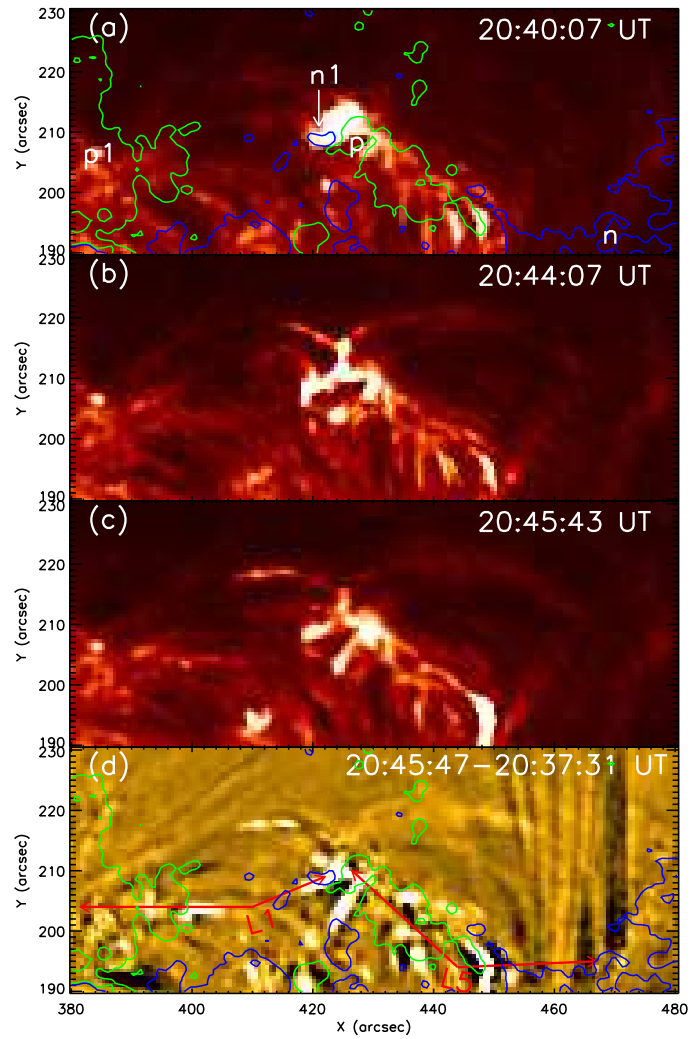}
\caption{AIA 304 \AA\ images ({\it a})\sbond({\it c}) and a 171 \AA\ difference image ({\it d}) showing the second loop-loop interaction process
that occurs between two sets of loops, L1 and ``L5" (as indicated by the red arrows). ``n" denotes a plage region where the negative footpoints of L5
are rooted. Iso-Gauss contours of  $\pm 100 G$ are also superposed by green and blue lines in panel ({\it a}) and panel ({\it d}).
An animation of this figure is available. The animation is 1s in duration, covering
20:37:19 UT to 20:54:59 UT.
(An animation of this figure is available.)}
\end{figure}

\begin{figure}
\epsscale{1.}
\plotone{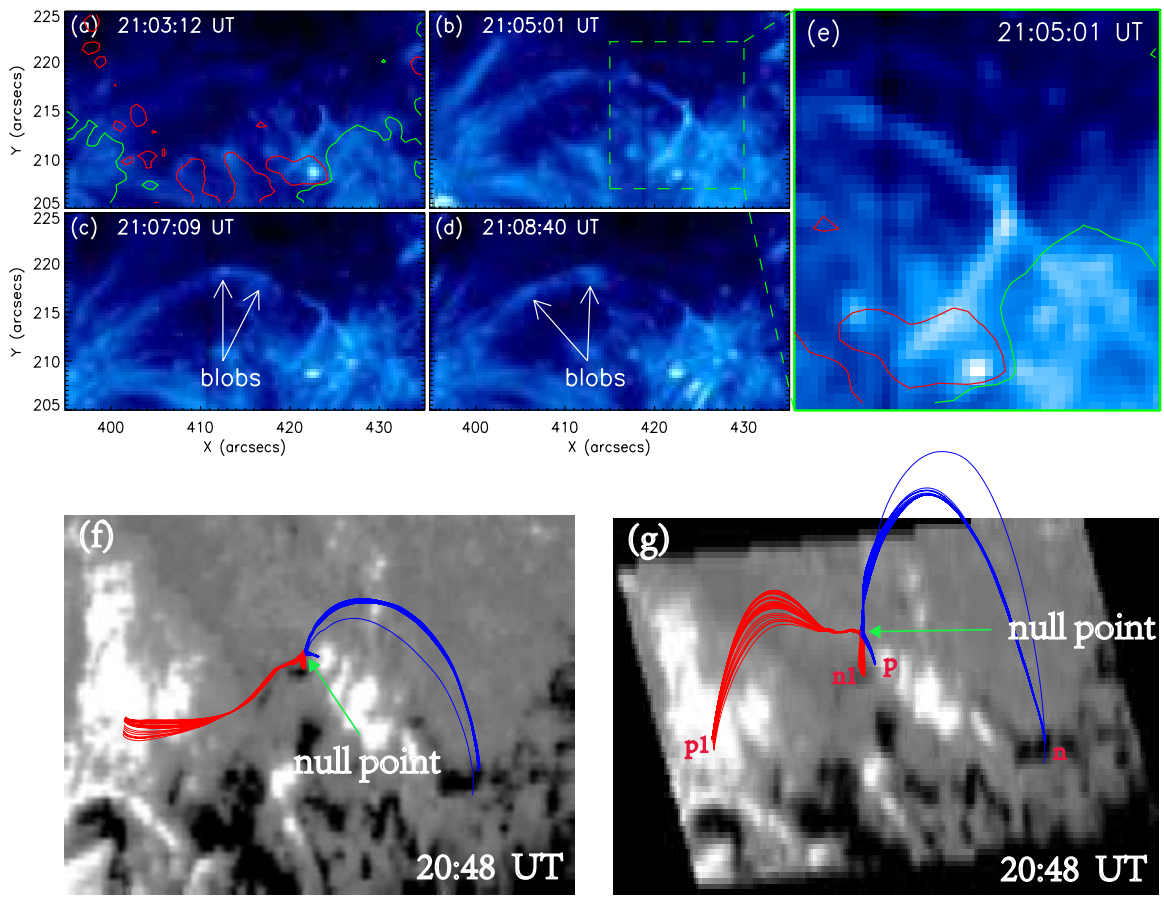}
\caption{Sequence of \emph{IRIS} 1400 \AA \ SJI images ({\it a})\sbond({\it d}) exhibit the successive ejection of plasma blobs from
a sheet-like structure above the flux cancellation sites. ({\it e}) A zoomed view corresponding to the green box in ({\it b}).
Iso-Gauss contours of  $\pm 30 G$ are also superposed by green and blue lines in panel ({\it a}) and panel ({\it e}).
NLFFF magnetic field extrapolation showing a top view (panel ({\it f})) and a side view (panel ({\it g})) of the coronal magnetic field
over the flux cancellation region at 20:48 UT. The red and blue lines represent field lines rooted in the cancelling
flux patches n1 and p, respectively. A magnetic null point (indicated by green arrows) lies above the flux cancellation sites.
It is located between the red and blue field lines. The background images are HMI vertical field images.
An animation of panels ({\it a}\sbond{\it e}) is available. The animation is 3s in duration, covering
20:59:51 UT to 21:15:08 UT.
(An animation of this figure is available.)}
\end{figure}


\end{document}